\documentclass[conference]{IEEEtran}
\IEEEoverridecommandlockouts
\usepackage{cite}
\usepackage{amsmath,amssymb,amsfonts}
\usepackage[normalem]{ulem}

\usepackage{algpseudocode}
\usepackage[hyphens]{url}
\usepackage{hyperref}
\usepackage[hyphenbreaks]{breakurl}
\usepackage{algorithm}
\usepackage{array}
\usepackage{graphicx}
\usepackage{textcomp}
\usepackage{xcolor}
\usepackage{svg}
\usepackage{hyperref}
\usepackage{boldline}
\usepackage{xspace}
\usepackage{multirow}
\usepackage{caption}
\usepackage{subcaption}
\usepackage{pifont}

\newcommand{\newComment}[1]{\textit{\# #1}}

\newcommand{\heading}[1]{\textbf{#1}}

\def\BibTeX{{\rm B\kern-.05em{\sc i\kern-.025em b}\kern-.08em
    T\kern-.1667em\lower.7ex\hbox{E}\kern-.125emX}}

\newcommand{\vaultor}{\texttt{VaulTor}\xspace}

\begin{document}

\title{VaulTor: Putting the TEE in Tor}

\author{

\IEEEauthorblockN{Humza Ikram}
\IEEEauthorblockA{
\textit{Carnegie Mellon University}\\
Pittsburgh, USA \\
humzai@andrew.cmu.edu}

\and 

\IEEEauthorblockN{Rumaisa Habib}
\IEEEauthorblockA{
\textit{Stanford University}\\
Stanford, USA \\
rumaisa@cs.stanford.edu}

\and 

\IEEEauthorblockN{Muaz Ali}
\IEEEauthorblockA{
\textit{University of Arizona}\\
Tucson, USA \\
muaz@arizona.edu}

\and 

\IEEEauthorblockN{Zartash Afzal Uzmi}
\IEEEauthorblockA{
\textit{LUMS University}\\
Lahore, Pakistan \\
zartash@lums.edu.pk}

}

\maketitle
\begin{abstract}

Online services that desire to operate anonymously routinely host themselves as `Hidden Services' in the Tor network. However, these services are frequently threatened by deanonymization attacks, whereby their IP address and location may be inferred by the authorities. We present \vaultor, a novel architecture for the Tor network to ensure an extra layer of security for the Hidden Services against deanonymization attacks. In this new architecture, a volunteer (vault) is incentivized to host the web application content on behalf of the Hidden Service. The vault runs the hosted application in a Trusted Execution Environment (TEE) and becomes the point of contact for interested clients. This setup can substantially 
reduce the uptime requirement of the original Hidden Service provider and hence significantly decrease the chance of deanonymization attacks against them. We also show that the \vaultor architecture does not cause any noticeable performance degradation in accessing the hosted content (the performance degradation ranges from 2.6-5.5\%).

\end{abstract}
\section{Introduction} 

The enormous expansion of the world wide web is coupled with growing demands for anonymity and privacy. Besides a huge end-user client base, an increasing number of web services---legal and illegal---also choose to remain anonymous~\cite{OnionServices}, fearing closure, or even prosecution, by the government and law enforcement agencies~\cite{onymous}. The Onion Router (Tor)~\cite{tor} network has emerged as one of the most popular solutions for providing anonymity: nearly 3 million clients connect to Tor daily~\cite{userMetrics}, and hundreds of thousands of anonymized web addresses are published each day.

Tor Hidden Services (aka. Onion services) aim to uphold freedom of speech in repressive regimes and offer circumvention 
in regions of undue and excessive internet censorship, thus bringing benefits to the public~\cite{good_tor}. At the same time, Hidden Services pave the way for criminal activities such as selling illegal drugs and weapons~\cite{silkroad}. All in all, there are incentives for governments and law enforcement to deanonymize Hidden Service (HS) providers\footnote{Hidden Service Provider is an individual that has the ownership of the Hidden Service. They also create and serve the content of the Hidden Service.} and curb their operations. In 2014, the authorities of 6 European countries and the United States collaborated in \textit{Operation Onymous} through which they cracked down on 416 Hidden Services, accused them of foul play, and arrested 17 individuals~\cite{onymous}.

Without arguing the legality and ethics of the Hidden Services, we address the technical challenges in reinforcing the anonymity of the HS providers over the Tor network. With this major goal in view, we present \vaultor, a novel overlay architecture for the Tor network, which builds upon the existing HS architecture by introducing a \textit{vault} overlay node between the HS provider and the clients. The HS provider offloads their service to the vault which now serves the clients on behalf of the HS provider. This setup strengthens the anonymity of HS providers in three key ways:

\begin{enumerate}
    \item After offloading their service, the HS provider does not need to be online all the time and may sporadically connect to the vault as and when needed (for any updates to the content). This significantly reduces the attack surface against long-running deanonymization attacks on the HS provider. 
    \item The HS provider may connect to the vault from widely varying locations, further reducing deanonymization risk. 
    \item An adversary cannot make an on-demand connection with an HS provider. Communication opportunities with the HS provider are now solely available to the vault, and even those channels are initiated by the HS provider.

\end{enumerate}

The service offloaded by the HS provider to the vault is hosted inside a Trusted Execution Environment (TEE), such as the one provided by Intel-SGX~\cite{sgx_intel}. Using a TEE brings two additional benefits: (i) the vault owner (even with root access) cannot snoop on the application code/data while it is being uploaded, served, or stored in the vault, and (ii) since client and the HS provider contact the vault through similar encrypted channels, the vault is never sure if it is an HS provider or a client it is communicating with, ruling out a myriad of attacks against the HS provider (details in Sec~\ref{sec:known-attacks-mitigated}).

The \vaultor architecture, which has no bearing on the client anonymity (discussed in Sec~\ref{subsubsec:client-privacy}), promises enhanced anonymity for the HS provider by shifting the deanonymization risk to the vault owner (see Sec~\ref{subsubsec:vault-privacy} for details). This design choice is suitable as the HS providers may only be interested in content creation without taking the deanonymization risk that leaps up when serving the content, even as a Tor Hidden Service. In contrast, the vault owners are willing to serve the content in exchange for monetary benefits (such as cryptocurrency payments\footnote{We discuss a secure way to do crypto transactions in section \ref{subsec:incentives}}) or social incentives (serving content they wish to support and propagate). A motivating example for our scenario might be a journalist in an oppressive regime who can anonymously upload content banned in their regime to a vault located in a neutral country. Another example may be providing `on-the-ground' information from within a region where Internet outages frequently occur (and the HS provider cannot remain online for extended periods of time).

\vaultor offers various attractive features not ubiquitous in alternate design choices; \vaultor (a) allows for hosting dynamic services, in contrast to data hosting services such as IPFS or \texttt{pastebin.com}, (b) offers content isolation from the server administrator, in contrast to a simple (non-TEE based) virtual machine-based approach, and (c) ensures that the server administrator cannot snoop on private keys in order to serve arbitrary content on behalf of the HS provider. 

\vaultor also brings robustness to the HS operation. If a vault node is shut down by the action of authorities, the HS provider can use a different one to make the content available. Alternatively, if the HS provider goes offline permanently, the TEE can continue hosting the content
as long as the vault owner remains incentivized.

In \vaultor, the vault simply replaces and acts on behalf of the HS provider, responding to clients through network paths containing the same number of Tor nodes as if the content is served by the HS provider in the existing architecture. Thus, the \vaultor design bears no negative impact on the network latency. We do observe a slight average increase in the time to first byte and time to last byte (a maximum of \textcolor{black}{$5.7\%$} and \textcolor{black}{$2.9\%$} respectively when using Intel-SGX as the TEE), attributed to hosting the content inside a TEE which incurs its own performance overhead (\ref{sec:results}).

To understand the operation of \vaultor, we note that a typical Tor circuit is an overlay path through three volunteer relay nodes (the entry guard node, a middle node and the exit node) on the Tor network~\cite{tor-og}. Such a circuit provides \emph{one-way anonymity} to a client trying to connect via the circuit to any non-anonymous server on the web. If the server wants to remain anonymous as well, it must also choose another Tor circuit with three nodes~\cite{trawling-for-hs}. The existing HS architecture anonymizes both the client and the server (two-way anonymity) using a mechanism that stitches the two Tor circuits together (see Sec~\ref{sec:current-architecture}). 
With the \vaultor architecture, a server that desires to be anonymous is facilitated to replicate, and infrequently update, its web content, and service on the vault. This is similar to a CDN node offering content replication to a server.
The two-way anonymity continues to exist between the vault node and the client, delivering similar delay performance as observed in the existing hidden service architecture. Indeed, our results in Sec~\ref{sec:results} confirm this. We also consider a sample of popular deanonymization attacks and provide an outline of how \vaultor offers enhanced HS provider anonymity under those attacks (see Sec~\ref{sec:known-attacks-mitigated}). Altogether, this paper makes the following contributions:
\begin{itemize}
    \item A novel architecture of Tor nodes, used by \vaultor to provide robust anonymity to host Hidden Services (HS).
    \item A step-wise description of the protocol \vaultor utilizes to ensure anonymity for the HS provider, vault, and the clients of hidden services.
    \item A thorough security analysis and description of the attacks that are mitigated by \vaultor.
    \item A working prototype (available at the anonymized link \url{https://anonymous.4open.science/r/vaultor-F197}) and its performance measurement over the actual Tor network.
\end{itemize}
\section{Background}

\subsection{Trusted Execution Environments (TEEs)}
\label{subsec:tee}
TEEs provide a platform for \emph{secure remote computation} which allows \emph{securely} executing an application in a remote \emph{untrusted} system without compromising the \emph{Integrity} and \emph{Confidentiality} of the application data. Several hardware architectures provide TEE implementations~\cite{sgx_intel,amd_sev,arm_trust,keystone}.
Our prototype implementation leverages the TEE provided by Intel-SGX~\cite{sgx_intel} to create a secure execution channel between the vault and the HS provider. We host an HS inside the TEE which itself is set-up inside a vault (Sec~\ref{valtor-sec}). An instance of a program running inside a TEE is called an enclave.

TEEs provide many guarantees. These include:
\begin{enumerate}
    \item \textbf{Sealing:} A program running inside a  TEE can encrypt and write data to the disk for persistent storage. Only the \textit{same program} running on \textit{the same device} in a TEE can decrypt this data.
    \item \textbf{Isolation:} A program inside a TEE can not access the memory of its host and vice versa.
    \item \textbf{Remote Attestation:} A piece of code running inside a TEE can prove, to an outside observer, what piece of code it is and that it is running inside a TEE.
\end{enumerate}

A local trusted device may use the following \emph{simplified flow} for remote attestation to verify that a remote untrusted device is running the desired piece of code:

\begin{enumerate}
    \item The local device builds the code and gets a measurement of the memory space\footnote{This includes the code itself, the data, the stack, and the heap.}.
    \item The local device sends the code to the remote device.
    \item The remote device runs this code inside a TEE. 
    \item When running inside a TEE, the code can then request the CPU to generate a cryptographic hash of the program's memory space. This hash is then signed by the CPU using a secret hardware attestation key embedded in the CPU. 
    
    \item The local device can request this signed measurement from the remote device. It then verifies this signature by matching the measurement hash against one computed locally and by verifying the signature against its public key.
    
\end{enumerate}

We will stick to Intel-SGX terminology, in which the signed cryptographic hash is called a \textit{quote}. Furthermore, it should be noted that a small amount of arbitrary data (produced by the code inside the TEE) can be embedded in the quote as well. This data is called the REPORTDATA field.
Different TEE implementations employ different means for signature verification. Intel-SGX, for example, requires sending the quote to Intel's online Attestation Service which verifies that the quote is valid. The local device also has the option to perform this attestation via a proxy (such as Tor).

The REPORTDATA field is crucial for establishing a secure connection with an enclave (program running within the TEE). This field routinely contains a public key whose corresponding private key is known only to the enclave. This public key can then be used to create a secure connection with the enclave using any form of key exchange such as Diffie-Hellman.

\subsection{Conventional Tor architecture} 
\label{sec:current-architecture}

In the current Tor protocol, there are 6 Tor nodes between a client and a hidden service (HS). This ensures two-way anonymity, i.e., 
both the client accessing the HS and the HS provider itself remain anonymous. Hidden services are identified using an onion URL.  

\begin{figure}
    \centering
    \includegraphics[width=\linewidth]{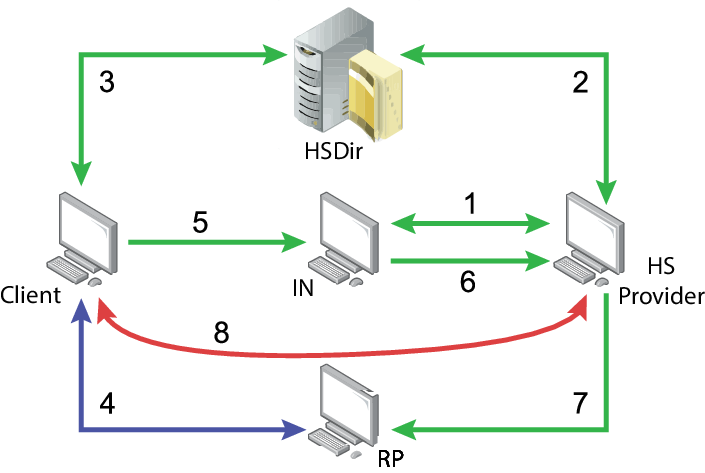}
    \caption{The current implementation of hidden services. Red paths represent information flow through Tor circuits with 6 nodes. Green paths represent information flow through Tor circuits with three nodes. Blue paths represent information flow through Tor circuits with only two nodes.}
    \label{fig:old_arch}
\end{figure}

Figure~\ref{fig:old_arch} and the following points detail the protocol for the establishment of communication between an HS and a client in the current architecture.
\begin{enumerate}
    \item An HS provider contacts a relay and asks them to act as an Introduction Point (IN)\footnote{IN is chosen to be distinct from Internet Protocol (IP)}. The HS provider receives an acknowledgment from the IN.
    \item The HS provider creates an ``Onion service descriptor'' which includes the public key for the HS and the IP addresses of its INs. This descriptor is signed by the public key of the HS. The HS provider sends this to a HSDir. The HS provider gets an acknowledgement from one of many Hidden Service Directories (HSDirs)\footnote{HSDirs are special relays that store and provide hidden service descriptors to the clients.}.
    \item A client asks the HSDir for the service descriptor of the HS. They receive and verify the signature of the HS.
    \item The client picks a Tor relay to act as a Rendezvous Point (RP) and establishes a Tor circuit to it. The client gives the RP a Rendezvous cookie.
    \item The client sends the same Rendezvous cookie and the IP address of the RP to an IN.
    \item The IN forwards the cookie and Rendezvous address to the HS provider.
    \item The HS provider makes a Tor circuit with the Rendezvous Point and sends the cookie. 
    \item The Rendezvous Point compares the two cookies and, if they match, relays communication from both sides to each other.
\end{enumerate}

\section{Related work}
There are two notable prior works that leverage TEE technology to improve Tor. 

In SGX-Tor~\cite{sgx-tor}, different nodes (such as Directory authority, HSDir, and middle nodes) run the Tor software in TEEs so that they cannot be modified to collude and launch deanonymization attacks.
In contrast, \vaultor reduces the deanonymization threats by not requiring the HS provider to be online.
Furthermore, \vaultor is an architectural solution that does not require any Tor node to run their service within TEE (unless they volunteer to be a vault), thus maintaining backward compatibility for Tor nodes.

Another approach, SmarTor~\cite{smartor}, utilizes TEEs and smart contracts to decentralize directory authorities (which are currently a few trusted and centralized servers) in the Tor network. The authors argue that this would increase the security of the Tor network as it would reduce the need to trust particular servers and make it more difficult for governmental authorities to crack them down. \vaultor, on the other hand, leverages the privacy guarantees provided by the TEEs to create trust by allowing a separate, new entity (the vault) to host web application content, thus making it harder to deanonymize the Hidden Service provider. 

Additional prior works (such as~\cite{vanguard,DeNASA,as-mit1,as-mit2}) have also suggested modifications to the Tor network to enhance HS anonymity. 
These works do not address the HS provider uptime requirement, but are fully compatible with \vaultor.

We discuss these solutions and their potential use cases in Sec~\ref{sec:enhanced-security}.

There exists a whole body of work aiming to improve the Tor network in general. Security improvements include a layered mesh topology~\cite{layered-mesh} for circuit formation and prevention of route capture attacks~\cite{shadowwalker}. There has also been work to improve client side stability in censored regions using WebRTC~\cite{snowflake, stegozoa,pokinghole}. These works do not address the uptime requirements for the HS provider.

Performance improvements include optimizing the Tor path selection algorithm \cite{lastor,navigator,path-claps,path-cong} and load balancing techniques~\cite{conflux, load-mleflow, load-peerflow}. These improvements can be applied in conjunction with \vaultor to further optimize the anonymity network (see Sec~\ref{sec:enhanced-security}).
\section{DESIGN GOALS} \label{sec:goals}

Our design goals are meant to ensure enhanced anonymity for HS-provider, while ensuring that anonymity guarantees for other parties (i.e., client, vault node) remain intact. 

We will outline our design goals succinctly here:
\begin{enumerate}
    \item Ensure that the anonymity guarantees for the HS provider in \vaultor are \textit{at least} as strong as an HS provider in the conventional Tor architecture. 
    \item Minimize the uptime for the HS provider, shrinking the attack surface against them.
    \item Ensure that the anonymity guarantees for the newly introduced Vault node are \textit{at least} as strong as those for an HS provider in the conventional Tor architecture.
    \item Ensure that the anonymity guarantees for the client remain \textit{at least} as strong as those for a client in the conventional Tor architecture.
    \item Ensure that the HS provider retains full control of any content that is being served on its behalf.
    \item Ensure that the system is able to serve rich content (not just limited to static content).
    \item Ensure that the system is backward compatible and is built using off-the-shelf components and technology.
\end{enumerate}

\section{Alternate hosting options}



The \vaultor architecture meets all our design goals specified in \S~\ref{sec:goals}. Other simple alternatives for anonymous content hosting may not satisfy this requirement.
This section considers two such possible alternatives and describes why they are unable to achieve the full set of our design goals.



\subsection{Using a static content hosting service}
An HS provider may upload static content to a data hosting service like an IPFS. While it is possible that the HS provider is able to upload content securely, this content can not be modified dynamically and, thus, fails to meet our fifth design goal. For example, an HS provider trying to a run a forum online will not be able to get posts or comments by clients. \vaultor permits dynamic content to be served which allows the HS provider to host complicated websites on an external server. 

\subsection{Hosting on the cloud}


An HS provider may host data dynamically on a virtual machine present on an external server (or a public cloud). They may use a remote login tool such as SSH (and the Tor proxy) to anonymously access the server and upload their content. However, the administrator of the server (with root access, or the cloud operator) may be able to access and modify this data even if the HS provider wishes to restrict access. This means the provider does not maintain full control over the content, violating the fourth design goal.

In \vaultor, the administrator cannot access data that is present inside the TEE. This data exists either in an encrypted manner on the disk or is only accessible to the TEE if the data is in the main memory. Lastly, as both the client and HS provider make an encrypted channel with the TEE, content is safe while in transit.

An HS provider that uploads their content to an external server (via SSH or through other means) is not guaranteed that this content will be served without modification. The administrator of this server may modify this content (while keeping the same URL and x509 certificate) and serve content that the administrator desires. This is possible because the administrator can snoop the private key from main memory or storage, presenting a major threat to client anonymity. On the other hand, \vaultor guarantees that the content served to a client is the content intended by the HS provider. The enclave creates a certificate and is the only entity (alongside the HS provider) that can serve this content. 

An HS provider may choose to provision cloud services that enable TEEs and host their content inside one. In this scenario, the cloud would act as an intermediary, reducing the uptime requirement for the HS provider. However, provisioning cloud services isn't anonymous. Know-your-customer (KYC) and/or payments via fiat currency can compromise the anonymity of the HS provider. It would also necessitate online transactions, which may be untenable in regions of instability, where volunteer vaults may be the only option. 






\section{Threat Model} \label{sec:thread-model}

We make realistic assumptions about the objectives and capabilities of adversaries. 
An adversary may have access to a small fraction of Tor relays and sufficient resources to qualify as a guard relay, HSDir, or a vault. Moreover, a malicious vault may be able to target specific HSes to host in their machine for the explicit purpose of deanonymizing the HS provider and clients of that service or for the purpose of modifying that service. 
Even a strong adversary will have a limited ability to acquire network traffic from ISPs and monitor traffic patterns between an end user and a guard node.  
We wish to protect the anonymity of the HS provider from this adversary and retain the control of the HS provider over any content that the HS is providing. 
In addition, a malicious HS may attempt to deanonymize a specific vault by hosting their web content on a TEE that the vault runs. As discussed in sec \ref{subsec:tee}, we do not consider side-channel attacks.

We also assume that any two entities (amongst the vault, HS provider and client) can collude to launch an attack against the third. For example, an HS provider and a client may work together in an attempt to deanonymize a vault.

Moreover, we assume a careful HS provider. That is, we assume that the HS provider does not leak identifying information in the content it provides. Furthermore, at time $t$, it will refuse to offload any content to a vault unless it verifies that, at time $t$, it is connecting to a valid enclave (this is a realistic assumption as verification is an easy task). Similarly, we assume a careful vault. That is, a vault inspects the program provided by an HS provider for malicious code, and will not run this program if it deems it malicious.

\section{VaulTor}
\label{valtor-sec}

\label{attacks-comparison}
\subsection{Architecture}
\begin{figure}
    \centering
    \includegraphics[width=\linewidth]{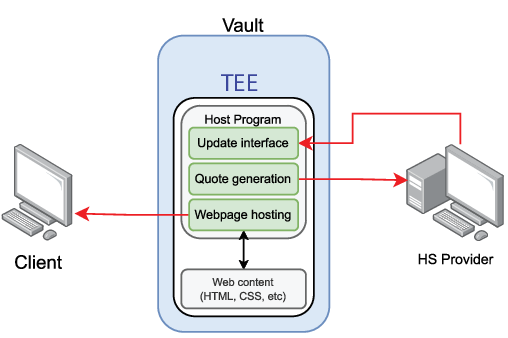}
    \caption{Our proposed implementation. Red paths represent information flow through Tor circuits with 6 nodes. The data inside the enclave is secure and information flow through the arrows is encrypted i.e., the vault owner can not interpret it.}
    \label{fig:Vault-arch}
\end{figure}

Building upon the conventional Tor hidden services architecture (Sec~\ref{sec:current-architecture}), \vaultor introduces three new entities: a device which we refer to as a \textit{vault}, a TEE which will host an \textit{enclave} and an optional external attestation service. The enclave 
is present inside the vault (see  Figure~\ref{fig:Vault-arch}) and utilizes its computational resources. 

In \vaultor, a willing device may offer to host content by advertising itself as a vault. This can be achieved without modifications to the current Tor architecture. The vault creates an onion address for itself (we shall refer to this onion website as the \textit{Vault Contact Hidden Service} (VCHS)) to facilitate correspondence with potential HS providers. Vaults may ask for compensation for hosting an HS (further discussion on incentives is given in Sec \ref{subsec:incentives}). In this scenario, the vault has the same privacy guarantees that the Hidden Services have in the current Tor architecture.

The HS provider can reach out to a vault through the vault's VCHS and provide a basic program which we shall refer to a \textit{host program} (HP) 
that 1) hosts a web server and 2) provides an interface through which (only) the HS provider can upload or remove content when they wish to update their website\footnote{The interface asks for a secret and after it is verified, the content can be uploaded or removed as desired. Since the secret verification occurs within the script that is running within an enclave, the vault cannot tamper with it without being noticed by the HS provider.}. This program must be running inside an enclave to provide the privacy guarantees detailed in this architecture. Thus, the HS provider must verify that the enclave has been correctly set up (i.e., the program is running without any modifications to the code and within a TEE) \textit{before} it provides all of the content it wishes to host and any incentives to the vault owner. Future content is also provided through the same update interface described earlier. This process is outlined in detail in Sec~\ref{boot}.

The TEE guarantees security and privacy for any content the HS provider transfers into the vault. In addition, it removes the requirement for continuous up-time of the original HS provider as the enclave can continue servicing the clients. 
While the vault must remain online to serve web content, the HS provider does not have this obligation and is hence protected from various deanonymization attacks (Sec~\ref{sec:attack_surfaces}).

Our architecture ensures increased anonymity and flexibility for the HS provider, minimal decrease in performance for a client (while maintaining the same security guarantees that Tor provides), and a level of anonymity for the vaults that is comparable to that of HSes in the current Tor architecture.

\subsection{Protocol}

\subsubsection{\textbf{Host Program (HP) Creation}} \label{subsec:protocol-hp}
Before an HS provider contacts a vault, it must write a host program that is intended to run within the TEE. This program should provide the following key functionalities:
\begin{itemize}
    \item It should host a server\footnote{Common servers such as Apache or Nginx can be used.} bound to a port. This server should be able to handle POST and GET requests. Furthermore, the code should be able to handle requests dynamically. For example, it should be able to store files uploaded by a client in separate directories.
    \item The HP should, by default, provide an interface on the server to input an authentication secret. If the secret (which is known to the HS provider) passes the hardcoded verification in the HP, the HS provider should be allowed to modify the contents of the enclave directory. The authentication key can either be a password that is hashed and compared, or it could be the HS provider's private key whose corresponding public key is written in the HP. 
    
    \item Upon starting the server, the program should generate a \textit{quote} file, that can be verified by the HS provider or a client (details given in Sec \ref{boot}). This quote file is available in the web directory and can be accessed by the HS provider or the client for verification through remote attestation as described in Sec \ref{subsec:tee}.
    
    \item The HP should provide some functionality to create backups of the hosted content in case the vault crashes. Backups are stored on the disk but encrypted using the enclave's \textit{sealing key}. This sealing key is deterministically generated by the CPU depending on the program running inside the TEE and the key burnt into the CPU. This key is only available to the enclave.
\end{itemize}

Running an arbitrary application inside a TEE is not straightforward. However, a library OS (libOS), such as Gramine-SGX, facilitates this process with minimal modifications to the application. Furthermore, since the entire libOS is contained inside the TEE, no inspection of the application code is necessary. To this end, we assume that a libOS like Gramine-SGX is available to the vault.

The pseudocode for a sample HP is given in Appendix~\ref{appendix:hp}.

\subsubsection{\textbf{Bootstrapping}} \label{boot}

The vault owner hosts and advertises the onion URL of its VCHS. The HS provider, vault owner and the HP (running inside a TEE) take the following steps to host their service in the vault (also shown in Figure~\ref{fig:new_arch}):

\begin{figure}
    \centering
    \includegraphics[width=1.0\linewidth]{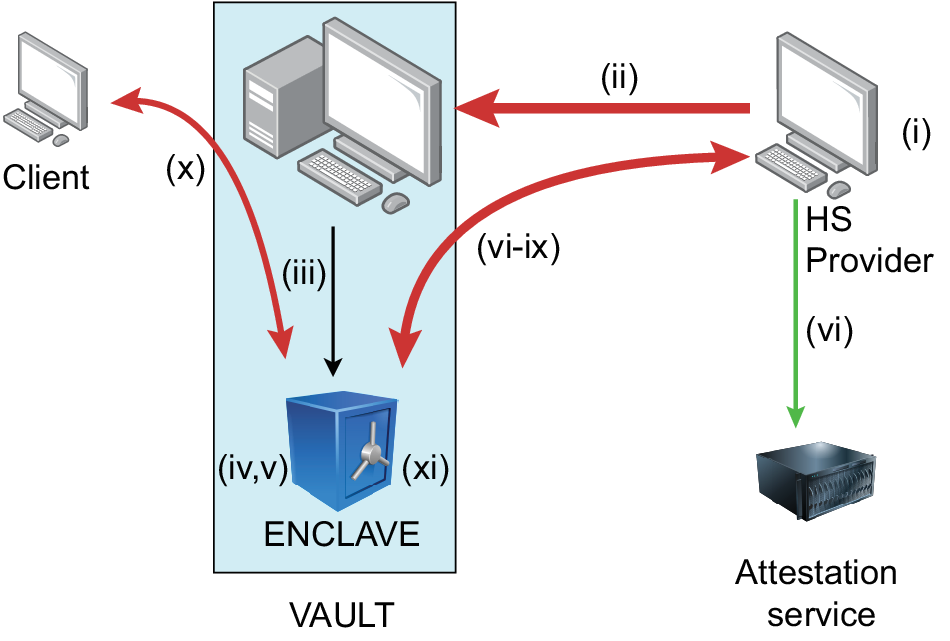}
    \caption{Our proposed implementation. Red paths represent 6-node circuits and green paths represent 3-node circuits.}
    \label{fig:new_arch}
\end{figure}

\begin{enumerate}
\renewcommand{\labelenumi}{(\roman{enumi})}
    \item The HS provider creates the HP (with the functionality described in Sec~\ref{subsec:protocol-hp}). 
   
    \item The HS provider uploads the HP at the VCHS. It is important to note that the host program is uploaded in plaintext (either as a script or a binary). 
    \item The vault owner runs the HP inside a TEE, hosts a hidden service (we shall refer to this as \texttt{yourHS.onion}), and binds it to the network port on which the host program will handle incoming requests.
    \item The host program will generate a public-private key pair ($P_{\mathrm{srv}}$, $S_{\mathrm{srv}}$) for the hidden service it provides. $P_{\mathrm{srv}}$ is made available to anyone who connects to \texttt{yourHS.onion} while $S_{\mathrm{srv}}$ only exists as a variable inside the enclave's memory (and is sealed to the disk for persistent storage). 
    \item The host program generates a quote which represents $P_{\mathrm{srv}}$\footnote{If the size of $P_{\mathrm{srv}}$ is too large to fit inside the REPORTDATA of the quote, a hash of $P_{\mathrm{srv}}$ may be used. If $P_{\mathrm{srv}}$ is embedded in a certificate, then typically the certificate hash is used.} and the program.
   
    This quote 
    is available to anyone accessing \texttt{yourHS.onion}.
    \item The HS provider accesses \texttt{yourHS.onion}, retrieves the quote, and  verifies that the quote is legitimate\footnote{In Intel-SGX, this may involve sending this quote to Intel's online attestation service. In RISCV Keystone, the end user can verify the quote themselves.}.
    \item The HS provider creates a secure connection with the enclave using $P_{\mathrm{srv}}$ 
    and any form of key-exchange such as Diffie-Hellman. This public key $(P_{\mathrm{srv}})$ may be embedded in a self-signed x509 certificate in order to facilitate \textit{https} connections.
    \item The HS provider supplies the authentication secret over this secure connection, after which it can securely upload content (such as HTML, CSS, PHP, and JavaScript files).  
    \item The enclave hosts the uploaded web application content.
    \item Any client that connects to \texttt{yourHS.onion} can interact with the hosted web application.
    \item The enclave regularly encrypts and backs up these files into non-volatile storage using the sealing key.
\end{enumerate}

\begin{figure*}
    \centering
    \includegraphics[width=0.85\linewidth]{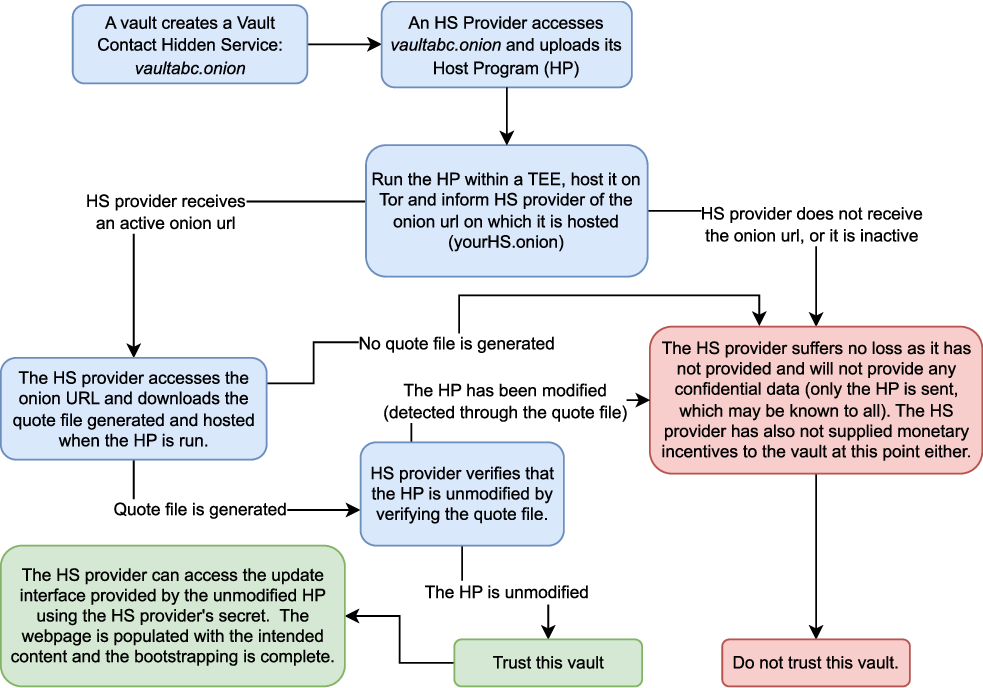}
    \caption{The steps taken in order for an HS provider to trust and upload to a vault.}
    \label{fig:trust-steps}
\end{figure*}

This procedure ensures increased anonymity for the HS provider. Their data is hosted in the vault and the vault owner can not access data inside the enclave or read the traffic in or out of the enclave. Figure~\ref{fig:trust-steps} provides a flow diagram of this process detailing some decisions the HS provider must make while uploading to the vault. 

\subsection{Enclave Isolation}
\label{subsec:enclave-isolation}
To ensure the protection of the vault, the enclave must have limited privileges. The following conditions, at minimum, are necessary:
\begin{enumerate}
    \item The enclave has access to a fixed, and limited, amount of RAM. One possible way to achieve this is by running the enclave within a virtual machine\footnote{The maximum ram must be set in Intel-SGX at enclave creation time.} configured with limited memory. This prevents an enclave from occupying all available RAM, thus safeguarding the performance of other programs on the vault.
    \item The enclave can only use a fixed amount of persistent memory. This can be achieved by isolating the enclave in a separate disk partition. This prevents an enclave from completely filling up persistent disk space that should be available to other programs on the vault.
    \item The enclave can only use a fixed amount of network resources. This is possible by controlling the network traffic rate through Trickle~\cite{trickle}. 
    \item Any connections going out from the enclave \textit{must} be restricted to only go via the Tor Proxy. This can be done by creating rules in iptables~\cite{iptables}. This is necessary to ensure that the IP of the vault is not made available to the HS provider. 
    Any traffic that is \textit{not} going out through port 9050 (the default Tor Proxy port) is blocked by a firewall.
    Furthermore, any traffic going to the Host Program must originate from the Tor client \footnote{This is to done to preserve vault anonymity from a malicious HS provider as discussed in \ref{subsubsec:vault-privacy}.}.
\end{enumerate}

\subsection{Client connection}
\label{subsec:client-connection}
A client must ensure that it is connected to the correct HS identified by its $P_{\mathrm{srv}}$ (embedded in an x509 certificate). To this end, the HS provider distributes not only the onion URL but also the hash of the x509 certificate when it wishes to advertise its service (similar to how conventional onion URLs are advertised). When connecting to an HS hosted on a vault, the client only needs to verify that the x509 certificate supplied by the service matches its advertised hash to ensure that it is connected to the appropriate entity. The client may maintain a list of valid certificate hashes\footnote{Checking of certificate hash is trivial and may be added as a subroutine in the client's Tor browser.}. Note that the client \textit{never} has to perform remote attestation themselves.

The client in \vaultor is exactly the same as a client in the conventional Tor architecture -- they use the same connection protocol. Layered on top of this is the ability to match an x509 certificate (or its hash) with the one that is advertised by the HS provider. This matching can be done trivially with a browser extension to the Tor browser. 

\label{subtable:attribute}

\section{Attack surfaces} \label{sec:attack_surfaces}

In this section, we specify the anonymity guarantees the \vaultor provides to each of the three entities: the client, the vault, and the HS provider. We consider the scenarios where each of these can be malicious as well as the scenario in which two of them collaborate to deanonymize the third. \vaultor enhances HS provider anonymity and leaves the client anonymity as it is. We further show that our new actor---the vault---is as protected as an HS provider in the current Tor Hidden Services architecture.

\subsection{{HS provider Anonymity}}
\label{subsubsec:hs-privacy} 
\ \

\heading{Scenario 1: Malicious client}

The HS provider no longer interacts with the clients directly. To access the web application content, the clients now establish a connection with the vault instead. Thus, unless the HS provider leaks identifying information in their content, they are safe from deanonymization at the hands of a client.

\heading{Scenario 2: Malicious vault}

In the traditional Tor design, where the client and HS provider communicate over a two-way anonymous channel initiated by the client, the attack scenarios in Table~\ref{table:attacks} render the client a harder anonymity target by a malicious HS provider than an honest HS provider by a malicious client. A mirror situation exists in the \vaultor design where the HS provider uploads and updates the content on a vault over a two-way anonymous channel. Thus, even in the worst case, an HS provider in \vaultor is as anonymous as an HS provider in the traditional architecture. Furthermore, the minimal uptime requirement enhances the anonymity of the HS provider in \vaultor architecture. The use of TEE at the vault offers additional guarantees of data integrity and data confidentiality to the HS provider.

\heading{Scenario 3: Vault and client collude}

The attack opportunities open to a client are a subset of the attack opportunities possible for a vault (since a vault has the same privileges as a client and more). Thus, the protection guaranteed for an HS provider from the vault applies in a scenario where the vault and client may collude.

\subsection{Client privacy}
\label{subsubsec:client-privacy}

We now show that a client is as protected in \vaultor as they are in the current Tor architecture.

\heading{Scenario 1: Malicious HS provider}

The HS provider is completely disconnected from the client, and hence is unable to launch attacks on the client directly.

\heading{Scenario 2: Malicious vault}

A malicious vault may attempt to a) serve modified content or b) launch a deanonymization attack on the clients. We now show why these attacks are not feasible in our architecture:

\begin{enumerate}
    \item[a)] 
    As the content is being hosted inside an enclave, clients can ensure that any content being served by the vault has not been maliciously modified. 
    Since the x509 certificate is generated by the HP running inside a TEE and the corresponding private key ($S_{\mathrm{srv}})$ is \textit{only} available to the TEE and the HS provider, a secure connection established using the certificate is guaranteed to be serving content vetted by the HS provider.
    Another consideration is that the content being hosted inside a vault is regularly encrypted and backed up to the disk. While this backed-up data can not be modified, a vault owner can selectively delete this backed-up data and restart the program in the enclave. This may result in the enclave accidentally serving outdated data to clients. However, if pieces of content are properly timestamped, the TEE can refuse to serve content that is outdated or add warnings while serving this content.
    \item[b)] In \vaultor, a client's perspective of the Hidden Service architecture remains the same. A client still accesses content through a 6-node connection--except that instead of connecting to the HS provider, it connects to a vault. Thus the client enjoys the same privacy guarantees as a client in the conventional Tor architecture. Furthermore, as discussed in sec \ref{subsec:client-data-privacy}, the client may enjoy enhanced data privacy.

\end{enumerate}

\heading{Scenario 3: Vault and HS provider collude}

A colluding vault and HS provider in the \vaultor architecture have the same attack opportunities against a client as a malicious HS provider in the conventional Tor architecture. Thus, the deanonymization risk for the client is the same as when only the vault is malicious.

\subsection{Vault privacy}
\label{subsubsec:vault-privacy}

\vaultor introduces a new entity in the Tor architecture: a vault that assumes a role similar to that of an HS in the current Tor design, thus maintaining a similar level of protection against deanonymization attacks.

\heading{Scenario 1: Malicious client} 

A vault is as vulnerable to a client as a Hidden Service is in the traditional Tor network.
One may even argue that the vault has stronger anonymity guarantees 
due to the fact that the web application content is hosted within an enclave (a secret, protected environment the vault cannot modify). This may provide plausible deniability to the vault as it would be blind to the traffic that enters and leaves its enclave.

\heading{Scenario 2: Malicious HS provider}

A vault is protected from attacks launched by an HS provider through the Host Program \textcolor{black}{by ensuring that the safety criteria specified in Sec \ref{subsec:enclave-isolation} are satisfied.}

Consider a malicious HS provider who uploads code that tries to obtain identifying information about the vault by leveraging the fact that the enclave is utilizing the vault's hardware. For example, this code may try to read files belonging to the vault or the vault's OS in order to \textit{directly} find identifying information or it may try to obtain its IP \textit{indirectly} by pinging an external server. The security guarantees provided by TEEs make direct attempts impossible; the host is \textit{also} isolated from the TEE just as the TEE is isolated from the host. Furthermore, most applications for TEEs run in a VM (as is the default in  gramine-SGX~\cite{gramine}), adding to the isolation. Indirect attacks are also mitigated using a firewall. By ensuring that all outgoing traffic is ported through port 9050 (the default port for Tor), only the IP of the exit node is available to the Host Program.

Furthermore, by ensuring that all requests originate from the Tor client software on the vault, the vault is protected from pinging based attacks. If this is not done, a malicious HS provider may upload a simple Host Program that replies with a unique phrase to the HS provider's IP when this Host Program is pinged. The HS provider may then ping various candidate IPs in the hopes of stumbling upon the vault's IP which would reply with the phrase. This sort of attack is only possible if the Host Program can be pinged directly, from outside the Tor client software.

\heading{Scenario 3: Client and HS Provider collude}

The HS provider is in a unique position as it directly provides code that the vault runs within an enclave. If an entity controls both the HS provider and a client node, we consider an attempt to launch a watermarking attack (described in Sec \ref{sec:known-attacks-mitigated}). This attack, in particular, only requires the control of two entities connected to the third. Moreover, the fact that the HS provider and client can make repeated on-demand requests to the vault further benefits the viability of this attack.

To attempt to launch a watermarking attack (similar to what~\cite{duster,wm_inflow} describe), the HS provider would add some watermark to the Tor traffic that can be identified at the client end. Despite controlling both the HS provider and a client, an attacker would not have the ability to successfully launch a watermarking attack on the vault. This is because of a missing component that this attack requires: the control of a guard relay. If we also assume the control of a guard relay, the HS provider is no longer necessary, as the guard relay can be the entity that watermarks the traffic. A guard relay and client could potentially launch this attack on their own without the requirement of an HS provider. Hence, the vault is as protected from a watermarking attack as an HS is in the current architecture. \textcolor{black}{Scenario 3 is thus akin to having two malicious clients in the conventional Tor architecture.}

\section{Evaluation}

In this section, we will qualitatively evaluate the effect of \vaultor in deflecting various families of existing attacks on HSes from the HS provider to our new Vault node. Afterwards, we will quantitatively measure the performance impact of \vaultor on client side network performance. 
\subsection{Known Attacks Deflected} 
\label{sec:known-attacks-mitigated}

 We will list various attacks and briefly explain how the \vaultor architecture deflects them from the HS provider to the vault. This list
 is non-exhaustive yet exemplifies the prominent attacks in recent years. 
 
\heading{Clock Skew:} 
     These attacks rely on repeatedly sending requests to an HS in order to heat up its CPU which has some tangible effect on the timestamp of incoming packets~\cite{clock-skew, clockskew2}. In \vaultor, this is impossible. \textit{No one} can repeatedly send packets to the HS provider.

     \heading{Congestion:} This attack relies on an adversary congesting existing guard nodes in the network~\cite{sniper-attack}, forcing the HS to connect to their compromised guard node long enough for the adversary to correlate traffic. This attack is completely deflected in \vaultor; the HS provider is sporadically online for limited periods and an adversary would have to congest the network indefinitely.

    \heading{Fingerprinting:} These attacks rely on learning the traffic patterns of an HS and referencing this against the traffic of a candidate set of guard nodes~\cite{devadas,circuit-finger,deepcorr,k-fp}. In our architecture, the vault is serving the traffic while the HS provider is taciturn. Thus, this type of attack will not work on the HS provider. Similar attacks that rely on compromised middle nodes \cite{insidejob} are similarly deflected.

    \heading{Guard Node Discovery:} 
    The Tor developers currently consider this the most potent threat against hidden services~\cite{vanguard}. This attack relies on making multiple connections with the HS provider such that their malicious middle node is next to the HS provider's guard node. Repeated, on-demand connections with the HS provider are impossible in \vaultor. As such, this attack is eliminated.
    
    \heading{Location Leaks:} Such attacks rely on the negligence of the HS provider and are out of the scope of this paper~\cite{caronte}.
   
    \heading{Watermarking:}
    In this type of attack, an adversary watermarks traffic on the client side in order to detect it at the malicious guard node of the HS provider~\cite{duster,wm_inflow}. 
      If a malicious vault node tries to launch this attack on the HS provider, this attack would be rendered less effective because HS-provider would have minimal uptime connection with the vault code instead of a constant connection.

Table~\ref{table:attacks} shows the various scenarios in which an adversary can launch a deanonymizing attack on Tor Hidden Services along with the impact of \vaultor on these attacks.

\begin{table}[h!]
\centering
\footnotesize
\begin{tabular}{lll}
\hlineB{2}
\multicolumn{1}{c}{Scenario} & \multicolumn{1}{c}{Attack Categories} & \vaultor impact \\ \hlineB{2}
\begin{tabular}[c]{@{}l@{}}Adversary can send\\ arbitrary requests\\to the HS provider\end{tabular} & \begin{tabular}[c]{@{}l@{}}Clock Skew, Watermarking, \\ Guard Node Discovery,\\ Fingerprinting\end{tabular} & \begin{tabular}[c]{@{}l@{}} Scenario \\ eliminated \end{tabular} \\ \hlineB{0.1}
\begin{tabular}[c]{@{}l@{}}HS Provider has \\ high uptime\end{tabular} & \begin{tabular}[c]{@{}l@{}}Clock Skew, Watermarking,  \\ Congestion, Fingerprinting\end{tabular} & \begin{tabular}[c]{@{}l@{}}Scenario \\ diminished\end{tabular}           \\ \hlineB{0.1}
\begin{tabular}[c]{@{}l@{}}High volume of \\ traffic coming from\\ the HS provider\end{tabular} & \begin{tabular}[c]{@{}l@{}}Watermarking,  Congestion,\\ Fingerprinting\end{tabular} & \begin{tabular}[c]{@{}l@{}}Scenario \\ diminished\end{tabular}           \\ \hlineB{2}
\end{tabular}
\caption{Various scenarios that lead to categories of contemporary attacks on Tor Hidden Services along with the impact of \vaultor on these scenarios.}
\label{table:attacks}
\end{table}

\begin{figure*}
\begin{subfigure}{0.32\linewidth}
\centering
    \includegraphics[width=0.95\textwidth]{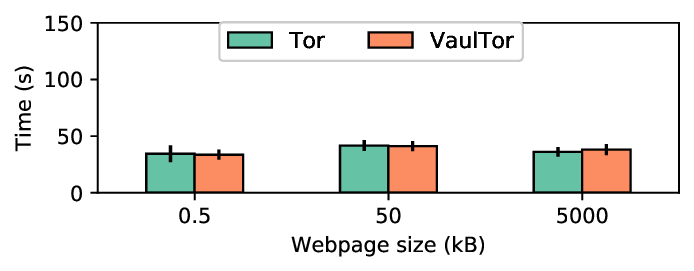}
    \subcaption{Random Relays, TTFB}
\end{subfigure}
\begin{subfigure}{0.32\linewidth}
\centering
    \includegraphics[width=0.95\textwidth]{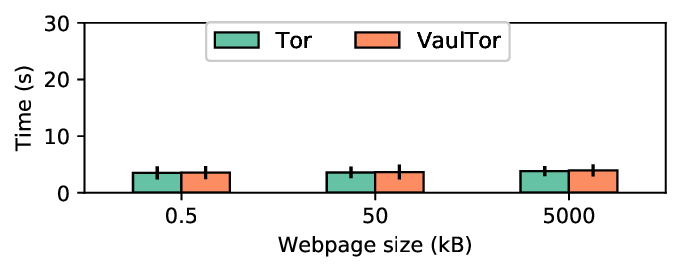}
    \subcaption{Fixed Relays, TTFB}
\end{subfigure}
\begin{subfigure}{0.32\linewidth}
\centering
    \includegraphics[width=0.95\textwidth]{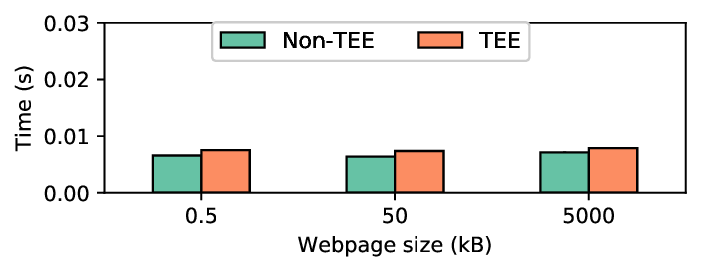}
    \subcaption{Local, TTFB}
\end{subfigure}
\begin{subfigure}{0.32\linewidth}
\centering
    \includegraphics[width=0.95\textwidth]{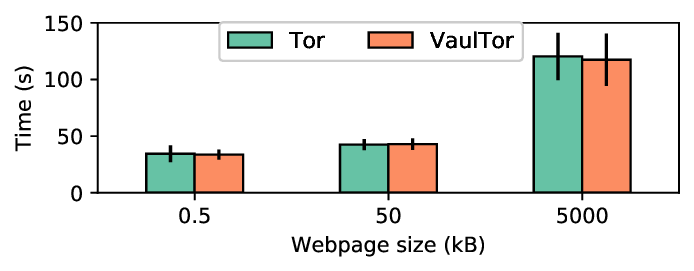}
    \subcaption{Random Relays, TTLB}
\end{subfigure}
\begin{subfigure}{0.32\linewidth}
\centering
    \includegraphics[width=0.95\textwidth]{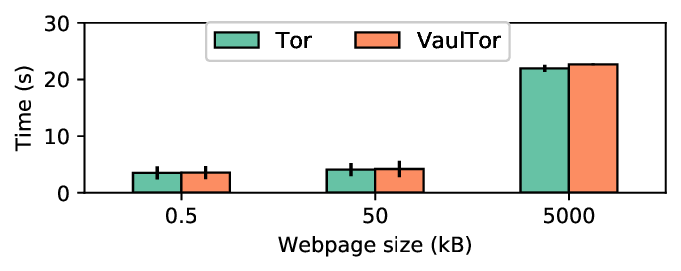}
    \label{subfig:fixed-ttlb}
    \subcaption{Fixed Relays, TTLB}
\end{subfigure}
\begin{subfigure}{0.32\linewidth}
\centering
    \includegraphics[width=0.95\textwidth]{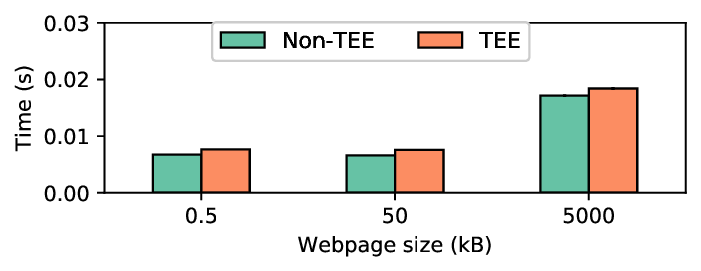}
    \subcaption{Local, TTLB}
\end{subfigure}
    \centering
    \caption{Time to first byte (TTFB) and time to last byte (TTLB) for webpages with varying page sizes without and within a TEE. Error bars represent 99\% confidence intervals.} 
    \label{fig:performance}
\end{figure*}

\subsection{Performance}

It is important that the security improvements \vaultor brings do not significantly degrade client side network performance. 
Important client side performance metrics include the network latency and the throughput. 
In this section, we detail our experimental setup (which includes our implementation of a vault) and our evaluation of these metrics. Our experiments measure the time experienced by the client to retrieve the data from an HS; the registration and bootstrapping processes in \vaultor occur only once and have a negligible\footnote{This performance impact will only be negligible if uploading content does not significantly increase the uptime of HS provider.} performance impact in the overall lifetime of the HS. 

\subsubsection{Experimental setup}

To measure and compare the network performance of Hidden Services when hosted within and outside a TEE, we ran two instances of a Host Program on the same machine. One of these HPs ran inside a TEE (facilitated by the gramine-SGX library OS~\cite{gramine}) while the other HP (which we shall refer to as a \textit{vanilla} HP) ran outside a TEE.

A Tor client\footnote{This client is not the same as a client in Tor architecture which we have discussed above. This is a program necessary to interact with the Tor network.} was also launched on the same machine which generated two onion URLs: one for the enclave and one for the vanilla HP. The Tor client directs traffic for each of these onion URLs to their respective HPs, allowing them to serve content via Tor. 

Both the enclave and the vanilla HP ran webservers and, in order to ensure consistency, served the same landing webpage simultaneously. In addition, the HP running inside a TEE had the ability to generate a quote in order to facilitate remote attestation. Both web servers were written in Python3 and regularly backed up data to persistent memory. Moreover, these webpages were hosted on the same device with an SGX-enabled Intel processor (Core-i5 10210U).

We conducted these experiments with three webpages, each with varying page sizes (0.5kB, 50kB, and 5000kB). The content on these webpages included HTML and JavaScript. 

We measured the performance using three methods: 
\begin{enumerate}
    \item \textit{Random Relays:} We restarted the Tor application between each measurement to establish fresh circuits. This gave us three random relays for every measurement. 
    \item \textit{Fixed Relays:} We used fixed/constant relays\footnote{These were chosen randomly from advertised Tor relays here: \texttt{\url{https://www.dan.me.uk/tornodes}}}
    across webpages for both \vaultor and Tor. We report the average performance of three different fixed circuits. 
    \item \textit{Local:} We locally accessed the webpages. 
\end{enumerate}
Each webpage was loaded 250 times in each of the methods, save for the Fixed Relays method, for which we loaded each webpage 250 times on each circuit (a total of 750 measurements) and took the average of the results. Methods (1) and (2) were conducted on the actual Tor network.

We thus quantified any overall changes in performance caused specifically by hosting an HS within a TEE.

{\subsubsection{Results}
\label{sec:results}
Figure \ref{fig:performance} shows the time to first byte (TTFB) and time to last byte (TTLB) for the 3 webpages hosted in the 2 architectures (the current architecture and \vaultor).

We note a minimal difference in performance across webpages and testbeds. Note that, for most of the results collected over Tor, the average TTFB and TTLB in the \vaultor architecture fall within the 99\% confidence interval of those of the current architecture. The only result (collected over the Tor network) that lied outside the confidence interval was for the TTLB of a 5000kB webpage routed through fixed relays. This had an average increase of 2.9\%.

If we consider all the results, including those that lie within the confidence intervals, we note a maximum increase in TTFB and TTLB of 5.7\%
(5000kB, Random Relays) 
and 2.9\% 
(5000kB, Fixed Relays),
respectively.

It should be noted that running an arbitrary program inside an Intel-SGX TEE may have a non-negligible computation overhead. 
When accessing the webpages locally (and hence, not over the Tor network), we note a maximum percentage increase in time in the case of the TTFB of a webpage of size 50kB (15.9\%). However, this delay is negligible compared to delays caused by Tor's network latency.

As such, it is not surprising that the percentage performance overhead of \vaultor over the conventional Tor architecture is minimal when measured over the real Tor network. 

We believe this nominal decrease in performance is justified considering the major anonymity benefits \vaultor brings to the HS provider. 

\subsubsection{Ethics} 

We had ethical considerations while conducting our performance measurements. We did not collect any deanonymizing information about any Tor relays and our load on the Tor network was negligible.

\section{Discussion}

\label{sec:discussion}
\subsection{{Additional Anonymity Measures}} \label{sec:enhanced-security}

There exist a number of proposals (such as ~\cite{vanguard,DeNASA,as-mit1,as-mit2}) that enhance the anonymity of the HS provider. These solutions, however, result in degraded network performance (longer delays and lower throughput) for the client, when used in the conventional HS architecture. This reduction in network performance renders these solutions less attractive today. With the \vaultor architecture, the vault serves the content to the clients, and any retrofitting at the HS provider side has no bearing on the network performance experienced by the client.

For example, the Vanguard add-on \cite{vanguard} (which inserts additional hops in the connection) may be used by the HS provider without affecting client-side performance. Similarly, privacy-preserving path selection methods \cite{DeNASA,as-mit1,as-mit2} may incur latency costs but are a non-issue for an HS provider that connects with the vault infrequently.

Furthermore, techniques like temporary proxies are completely compatible with our system and may be used by clients and the HS provider to connect to vaults to obfuscate their traffic \cite{snowflake, stegozoa, pokinghole}.

\subsection{{HS Provider Flexibility}}
\label{hs-mobility}

In the current Tor architecture, the HS provider must remain static in order to serve content. The flexibility offered by \vaultor can be leveraged by the HS provider to communicate from secure and variable locations. This would especially be beneficial in the context of activists or journalists who want to report their content from secure \textit{intermediate} locations in oppressive regimes without the risk of getting caught. Moreover, \vaultor would allow the HS provider's content to remain accessible during Internet outages, which is commonplace in regions with political instability and censorship \cite{bangladesh-outage,summary-outage}. 

\subsection{Incentives and Smart Contracts} \label{subsec:incentives}
The vault owner proxies for the HS provider and, on its behalf, serves content to the clients. This act must be incentivized  
for the vault owner. These incentives may be social incentives -- similar to how users of Tor run relays and nodes.

However, if incentives are monetary, they must be exchanged in a secure and private manner.
Towards this end, a blockchain may be used to ensure that the vault owner receives cryptocurrency rewards for hosting content for the HS provider. One approach to this is that the vault owner supplies the HS provider with their address on a public blockchain such as Ethereum~\cite{ethereum}. Only if regular cryptocurrency payments are made to this public address does the vault owner continue hosting. This allows both the HS provider and the clients to ``crowdsource'' an HS on a vault.

This previous approach does necessitate timely payments from the HS provider. This requirement can be removed via the use of smart contracts. A smart contract can lock the cryptocurrency that it receives from the HS provider and clients. The smart contract can then use an oracle to verify that the HS is being hosted properly and perform remote attestation. If the HS is being hosted properly and the remote attestation is successful, the smart contract releases the cryptocurrency to the vault's blockchain address. In order to preserve privacy, zero knowledge enabled cryptocurrency such as Zcash~\cite{zcash-bib} can be used. Furthermore, a Decentralized Exchange (DEX) may be used to trade cryptocurrency. These measures reduce the possibility of profiling based attacks. 

For example, a vault owner may make their zero-knowledge blockchain address (such as for monero \cite{monero}) available on VCHS and the HS provider may anonymously transfer this cryptocurrency by publishing their transaction by connecting to an external blockchain node through the Tor proxy. In a zero-knowledge blockchain, the transaction itself will have no identifying information present in it that may be used for social engineering (such as, by using chain analysis tools). 

\subsection{Plausible Deniability for Vault}
In the \vaultor architecture, a vault owner is not privy to the content present inside the TEE.
We believe that this adds an extra layer of plausible deniability, greater than the plausible deniability of conventional data hosting services that are aware of the content being served.  
When hosting content on behalf of an HS provider, the only thing the vault owner knows is the Onion URL of the Hidden Service. This has a parallel with Guard Nodes in the conventional Tor architecture that know what Onion URL's traffic is routed through them. Both the vault and the guard node can not read this traffic or compromise its integrity, only help move this content. The only difference is that the physical storage resources of the Vault owner are being used. However, even this physical storage is encrypted and opaque to the Vault owner which is not privy to the information being served, just like a guard node.

In our future work, we can enable   vault owners  to run a Tor client inside an enclave and run the HP inside another enclave on the same machine. The Tor client generates a new onion URL and shares it (only) with the Host Program. The HP then serves content on this onion URL using the Tor client as a proxy. 
The HP also forwards the onion URL to the HS provider who can then advertise it as before. In this scenario, the client does not need to modify their browser.
As such, the vault owner is not aware of the content they are serving. In this scenario, they can not be held liable for the content they are serving as the information about which machine is serving what content is available to \emph{``no one''}. And \emph{``no one''} includes the vault owner and the HS provider.

\subsection{Incremental deployment:} The \vaultor architecture supports incremental deployment (albeit its strength is fully utilized when there are many vaults present on Tor). Vaults can register their VCHS themselves to an HSDir similar to how Hidden Services are currently already registered. This allows for a slow, optional adoption of \vaultor.

The client \textit{does} need to install a small extension (as discussed in Sec \ref{subsec:client-connection}) that compares $P_{\mathrm{srv}}$ (or its hash) with the one advertised by the HS provider but this is a trivial add-on and does not affect traditional HSes.

\subsection{Multiple Vaults:} An HS provider may commission multiple vaults to hold their data. To this end, they may download the $S_{\mathrm{srv}}$ and the x509 certificate from the HP of one vault and upload it to an HP they have hosted on another vault.
As clients use the certificate hash provided by an HS to validate its identity (as described in Sec \ref{subsec:client-connection}), they can be certain they are being served by the same HS provider even if the onion URL of the HS is different. This will add redundancy and fault tolerance to the HS provider's content: if one vault becomes inactive, the other vaults can continue to serve content.

\subsection{Strengthening client data privacy} \label{subsec:client-data-privacy}
In the \vaultor architecture, the host program is present inside a TEE and its measurement (see \S~\ref{subsec:tee}) is available to the client (in addition to the HS provider). The HS provider may elect to make the code itself available to the client, allowing the client to inspect this code. If the code is simple (for example, the code only stores and serves content to password authenticated requests), then the client can upload private data to the server \textit{without having to trust the HS provider} as is necessary in the current Tor architecture. 

\subsection{Attacks against TEEs:}A wide variety of side-channel attacks exist that can target TEEs~\cite{side-channel1,side-channel2,side-channel3,side-channel4,side-channel5}. These attacks aim to discern secrets contained inside the TEE, such as private keys, through various means such as leveraging page faults. Vendors are prompt in mitigating side-channel attacks~\cite{tee-fix5} as the community uncovers those. Considering the research and development that focuses on mitigating side-channel attacks~\cite{tee-fix1,tee-fix2,tee-fix3,tee-fix4}, architectural designs discount such attacks from their threat models~\cite{side-ignore1,side-ignore2,side-ignore3}. We also follow take the same course of action (\S~\ref{sec:thread-model}).

\subsection{Advancements/variations in TEE technology:} Although our current implementation utilizes Intel-SGX, \vaultor is a generic solution that could theoretically support any TEE service. As newer TEE services (such as Intel-TDX~\cite{intel-tdx}) emerge and improve, the strength and flexibility of \vaultor improve as well. In the future, a diverse set of vaults utilizing differing TEE technologies could be built and tested. Other promising implementations of TEEs are also being proposed~\cite{cure,sanctuary,sanctum,komodo,keystone}. 
We believe that TEEs will become increasingly resistant to side-channel attacks.

\section{Conclusion}
We present \vaultor as an architectural solution that leverages TEE technology to reduce the threat of deanonymization attacks against HS providers on the Tor network. To this end, \vaultor introduces a new actor: the vault, which serves content on the HS provider's behalf. We show that \vaultor prevents several HS deanonymization attacks by utilizing the vault, whilst preserving the same level of client anonymity as in the current architecture. This is achieved without any noticeable performance degradation experienced by the client. We also argue that vaults have the same security guarantees as HS providers in the conventional Tor architecture.

\bibliographystyle{IEEEtran}   
\bibliography{references}
\section{Appendix}
\subsection{Host Program}
Algorithm \ref{algo:hp} is an example of a simple HP written in a high level language. It should run inside a VM inside a TEE (which is the default case in Gramine-SGX).

\label{appendix:hp}
\begin{algorithm}
\caption{Host Program}
\label{algo:hp}
\begin{algorithmic} 

\State passwordHash = ``HardcodedByHSProvider''

\If{existsOnDisk($S_{srv}$) }
    \State $S_{srv}$ = fetchFromDisk(Spath)
\Else
    
    \newComment{Generate key pair using a TRNG as source}
    
    \State $P_{srv},S_{srv}$ = TRNGenKeyPair() 
\EndIf

\State sealToDisk($P_{srv}$, Ppath)
\State sealToDisk($S_{srv}$, Spath)

\noindent\newComment{Run a server and send all incoming requests to the HandleRequest function}

\State hostServer(``localhost'', port, $P_{srv}$, $S_{srv}$) 
\\
\Function{handleRequest}{request}
    \If{hash(request.password) == passwordHash} 
        \If {request.type == "quote"}
            \State writeReportData($P_{srv}$)
            
            \newComment{Get a quote signed by the processor}
            
            \State quote = getQuote() 
            \State return quote
        \ElsIf{request.type == ``upload"}
            \State modifyContent(request.data, request.path)
            \State return SUCCESS
        \ElsIf{request.type == ``download"}
        
        \newComment{Decrypt from disk and return}
            \State return fetchFromDisk(request.path) 
        \EndIf
    \ElsIf{request.owner == ``client"}
        \If{request.type == ``download"}
            \State return fetchFromDisk(request.path)
        \EndIf
    \EndIf
    
    \State return FAILURE
\EndFunction

\end{algorithmic}
\end{algorithm}

\end{document}